# In-plane epitaxy strain tuning intralayer and interlayer magnetic couplings in CrSe$_2$ and CrTe$_2$ mono- and bi-layers

Linlu Wu[1,†], Linwei Zhou[1,†], Xieyu Zhou[1], Cong Wang[1,*] and Wei Ji[1,*]

[1]*Beijing Key Laboratory of Optoelectronic Functional Materials & Micro-Nano Devices, Department of Physics, Renmin University of China, Beijing 100872, P.R. China*
Corresponding authors: C.W. (email: wcphys@ruc.edu.cn), W.J. (email: wji@ruc.edu.cn)
† These authors contributed equally to this work.

Mismatched lattice-constants at a van-der-Waals epitaxy interface often introduce in-plane strains to the lattice of the epitaxial layer, termed epitaxy-strain, where the strains do not follow the intra-layer Poisson's relation. Here, we presented the magnetic phase diagrams of the CrSe$_2$ and CrTe$_2$ mono- and bi-layers under epitaxy-strain up to 8 %, as predicted using density-functional-theory calculations. They show that the in-plane epitaxy-strain manipulates either the intra- or inter-layer magnetism of them. The in-plane strain varies the interlayer distance, defined using an inter-layer Poisson's ratio, which governs the inter-layer magnetism in two opposite ways depending on the in-plane magnetism. The tunability of the intra-layer magnetism is a result of competing intra-layer Cr-Cr super-exchange interactions. A graphene substrate was introduced to examine the validity of our diagrams in practice. Our work also gives a tentative explanation on the controversially reported magnetizations in CrSe$_2$ and CrTe$_2$ epitaxial mono- or bi-layers under epitaxy-strains.

Magnetism in two dimensional (2D) van der Waals (vdW) materials has received extensive attention in recent years. Strain engineering appears an effective route to manipulate magnetism in 2D materials, which was experimentally realized through, e.g., hydrostatic pressure for $CrI_3$ [1-3] and uniaxial strain for $Fe_3GeTe_2$ [4] recently. In a more ideal case, theoretical calculations showed that a uniformly in-plane biaxial strain could tune the magnetic ground states of $CrSe_2$ and $CrTe_2$ monolayers, which is yet to be experimentally verified [5]. In these studies, strain induced lattice variations are either uniform (biaxial strain or hydrostatic pressure) or coupled with the Poisson's ratio of the material. Thus, few of previous studies have dealt with the case where in-plane strains are independently applied along with the two lattice vectors of a 2D magnet, which are usually occurred in epitaxy growth of layered materials.

Epitaxy of heterostructures usually introduces lateral interfacial strains because of lattice mismatch, which is a long-lasting way to maintain in-plane strain to the adlayers and is thus termed epitaxy strain [6-8]. VdW epitaxy refers to growth of 2D layers through vdW interactions on a dangling-bond-free substrate [9, 10]. Magnetic 2D layers in vdW heterostructures show a strong ability to bear with large lattice mismatches and thus significant in-plane strains [11-13]. A recent illustration of this ability lies in epitaxy of monolayer $CrTe_2$ on graphene where a 7 % compressive and a 4 % tensile epitaxy strain were applied along the two lattice vectors, in which a zig-zag anti-ferromagnetic (AFM) order was observed [12]. This is not an isolated example that a 6 % epitaxy tensile strain was applied in the both lattice directions of a $CrSe_2$ monolayer through epitaxy growth on $WSe_2$ where a weak ferromagnetic (FM) state was reported [13]. In comparison with the ABAB order predicted in the free-standing $CrSe_2$ monolayer, it implies a potential ability of epitaxy strain to tune in-plane magnetism of 2D magnets, which is yet to be fully unveiled. In-plane magnetism aside, an interesting question is subsequently arisen that whether in-plane strains could change inter-layer spin-exchange couplings.

In this work, we comprehensively considered the roles of epitaxy strain in tuning intra- and inter-plane magnetic couplings in $CrSe_2$ and $CrTe_2$ mono- and bi-layers using density functional theory calculations. The predicted magnetic phase diagram of the $CrSe_2$ ($CrTe_2$) monolayer shows that its intralayer magnetic ground-state is tunable among FM and three AFM orders within a 2.5 % (4.5 %) in-plane strain. This tunability is primarily realized by changing Cr-Se-Cr (Cr-Te-Cr) angles and thus the strength and type of Se (Te) mediated super-exchange interactions between adjacent Cr atoms. Moreover, the varying in-plane strain also affects interlayer Se-Se or Te-Te distances and thus changes the interlayer magnetism between ferromagnetic (FM) and AFM in two different manners [14]. A CrTe$_2$ (1L)/Graphene (2L) heterostructure model, recently prepared in an epitaxy experiment [12], was used to verify the reliability and feasibility of the phase diagrams.

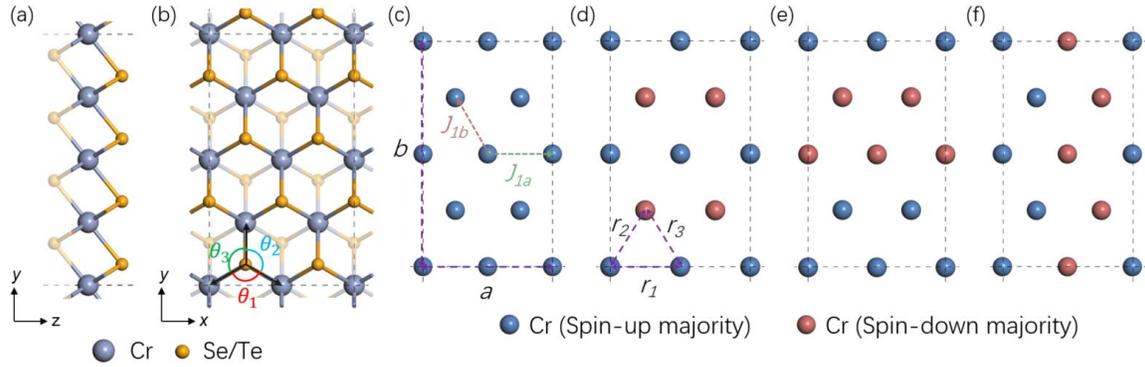

**Figure 1. Schematic models and magnetic configurations of monolayer CrX$_2$.** (a) and (b) Side and top views of monolayer 1T-CrX$_2$. The orange and grey balls represent top (solid) and bottom (semitransparent) layer Se (Te) and Cr atoms, respectively. Three colored arcs denote three Cr-Se-Cr (Cr-Te-Cr) angles $\theta_1$, $\theta_2$ and $\theta_3$, respectively. Grey dash rectangles label the $2 \times 2\sqrt{3}$ supercell used for considering four magnetic configurations in (c)-(f). The dark blue (red) balls indicate Cr atoms where spin-up (down) is the majority spin-component. Lattice constants $a$ and $b$ are labeled using two purple dash lines in (c). Green and brown dashed arrows in (c) denote intralayer spin-exchange parameters $J_{1a}$ and $J_{1b}$ (nearly) in the $a$ and $b$ directions between the nearest Cr sites, respectively. Distances $r_1$, $r_2$ and $r_3$ represent the nearest Cr-Cr distances in three lattice directions.

Both freestanding 1T-CrSe$_2$ and -CrTe$_2$ monolayers take a hexagonal crystal structure with the *P-3M*1 space group (Fig. 1a and 1b) in their paramagnetic states. A $2 \times 2\sqrt{3}$ rectangular supercell was used for considering four magnetic, i.e. FM (Fig. 1c), ABAB (Fig. 1d) and AABB (Fig. 1e) striped AFM (sAFM) and zigzag (ZZ, Fig. 1f) AFM, configurations in our calculations. Configuration sAFM-ABAB (ZZ-AFM) is the most two energetically favored configuration among all those considered ones of the CrSe$_2$ (CrTe$_2$) monolayer (see Table S1 for more details). The easy axis of CrSe$_2$ (CrTe$_2$) is oriented in the *y-z* plane and is 60° (100°) off the *z* axis (Fig. S1). Here, we employed the CrSe$_2$ monolayer as a prototype for discussion. Its fully relaxed FM structure (Fig. 1c), shows the lattice constants *a*=6.84 Å and *b*=11.84 Å, the nearest Cr-Cr distance $r_1$=$r_2$=$r_3$=3.42 Å and the corresponding Cr-Se-Cr angle $\theta_1=\theta_2=\theta_3=$ 84.6°, exhibiting a $C_3$ rotational symmetry, which was chosen to be the non-strained structure for further comparison.

The $C_3$ symmetry, however, breaks in those three AFM configurations. Introduction of AFM spin-exchange coupling shortens $r_2$ and $r_3$ to 3.32 Å but elongates $r_1$ to 3.50 Å. The values of those angles split in accordance with the changes of Cr-Cr distances that $\theta_2$ and $\theta_3$ decrease to 81.7° and $\theta_1$ increases to 86.8°. The lattice degeneracy further breaks in the ZZ configuration, which is the predicted groundstate for the $CrTe_2$ monolayer. In particular, the two AFM coupled distances $r_1$ and $r_3$ split into 3.59 and 3.57 Å, respectively, while the FM coupled distance reminds its FM configuration value of 3.66 Å. All associated angles decrease from 84.2$\pm$0.1° to $\theta_1$= 82.1°, $\theta_2$= 81.7° and $\theta_3$=83.9°. These results indicate that magnetic configurations in the $CrSe_2$ ($CrTe_2$) monolayer is tightly coupled with their geometric structures, i.e. lattice constants, $r_1$ to $r_3$ and $\theta_1$ to $\theta_3$. In light of this, it deserves a closer examination that the influence of geometric structures on magnetic orders in the both monolayers.

Figure 2a plots a magnetic phase diagram of the $CrSe_2$ monolayer as a function of lattice constants $a$ and $b$. Tensile and compressive strains up to 8% were applied to the fully relaxed FM structure, i.e. $a$=6.84 Å and $b$=11.84 Å. While the FM configuration favors in expanded lattice constants, compression in the $a$ ($b$) directions changes to the groundstate to the ZZ (ABAB) configuration. The AABB state appears to be the most stable in a very narrow window between the FM and ABAB phases. The phase diagram of $CrTe_2$ (Fig. 2b) shows a similar feature, but with much pronounced AABB region.

In the $CrSe_2$ ($CrTe_2$) monolayer, Se (Te) meditated Cr-Cr super-exchange interactions dominate its intralayer magnetism, which highly depends on the Cr-Se(Te)-Cr angle [15]. We thus defined spin-exchange parameters $J_{1a}$ and $J_{1b}$ along the two lattice directions (Fig. 1b) to explore the roles of varying in-plane epitaxy strain in changing $\theta_1$ (Fig. 2c), $\theta_2$ (Fig. 2e), $J_{1a}$ (Fig. 2d) and $J_{1b}$ (Fig. 2f) in the $CrSe_2$ monolayer as a prototype. Lattice constant $a$ directly affects angle $\theta_1$ and spin-exchange parameter $J_{1a}$, consequently the in-plane magnetic configuration. As shown in Fig. 2c-2f, in the FM region, angles $\theta_1$ and $\theta_2$ are very close to 90º and $J_{1a}$ and $J_{1b}$ are both negative in sign, which favors the Cr-Cr super-exchange. Angle $\theta_1$ is nearly independent of lattice parameter $b$ and gradually decreases with shrinking $a$ values, i.e. from 90º at $a$=7.30 Å to 76º at $a$=6.29 Å with fixed $b$=11.84 Å

(the strain-free constant $b$ value, Fig. 2c), which disfavors the Se meditated Cr-Cr FM super-exchange along the $a$ direction. As a consequence, $J_{1a}$ reverses its sign from negative to positive at, e.g. $a$=7.04 Å (~2.9 % tensile strain) with fixed $b$=11.84 Å, as marked using the purple dot in Fig. 2d, suggesting an FM to AFM transition. The sign reversal is the primary origin of the FM to ZZ-AFM transition.

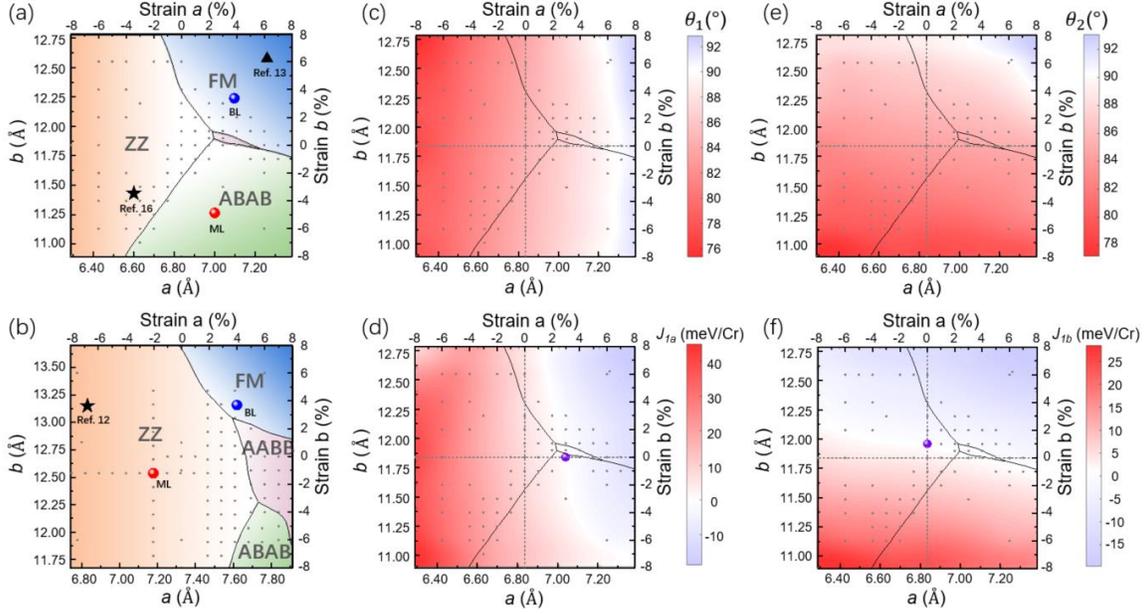

**Figure 2. Tunability of intralayer magnetism of monolayer CrSe$_2$ and CrTe$_2$.** (a) and (b) Phase diagrams of monolayer CrSe$_2$ (a) and CrTe$_2$ (b) as a function of lattice constants $a$ and $b$. Blue, orange, green and purple components represent intralayer FM, ZZ, ABAB and AABB, respectively. Each gray dot represents a theoretical data point that the energies of those four configurations are compared. Red and blue dots label fully relaxed lattice constants under the most energetically favorable magnetic configurations in mono- and bi-layer CrX$_2$ (X=Se, Te), respectively. Those black triangle and star symbols mark the lattice constants of experimentally synthesized monolayer CrX$_2$ (X=Se, Te) on different substrates. (c) and (e) $\theta_1$ and $\theta_2$ varying with the lattice constants in monolayer CrSe$_2$. (d) and (f) Mapping of exchange parameters $J_{1a}$ and $J_{1b}$ as a function of lattice constants. Positive (red) and negative (blue) areas represent AFM and FM spin-exchange coupling, respectively. Phase boundaries of monolayer CrSe$_2$ are labeled by black lines in (c) - (f).

The FM to ABAB transition shares a similar mechanism under shortening constant $b$. The shortened $b$ lattice, thus the decreasing $\theta_2$ angle (Fig. 2e), directly flips the sign of $J_{1b}$ at, e.g., $b$=11.96 Å (~1.0 % tensile strain) with $a$=6.84 Å (the strain-free constant $a$ value), as marked with the purple dot in Fig. 2f, but has little effect on $J_{1a}$ except with strongly compressed $b$ values. The flipped sign leads $J_{1b}$ to favor the AFM super-exchange at smaller $b$ values. Thus, the AFM coupling occurs in the $b$ direction, showing the sAFM-ABAB ground-state at the right bottom part of the diagram. Competition between the FM and AFM couplings along $b$ results in a FM-sAFM transition configuration, i.e. sAFM-AABB, appeared in a very small region in-between FM and sAFM-ABAB, in which the two types of magnetic super-exchange occur alternately. In terms of the boundary between ZZ-AFM and sAFM-ABAB, they show competing magnetic couplings in either $J_{1a}$ (AFM versus FM) and $J_{1b}$ (FM versus AFM). They thus roughly bisect the phase diagram outside the FM region.

The phase diagram of monolayer $CrTe_2$ (Fig. 2b) show a qualitatively similar but quantitatively different pattern to that of the $CrSe_2$ monolayer. While it is also comprised of those four phases, the ZZ-AFM phase occupies a larger region and the sAFM-AABB phase is more pronounced. A larger tensile strain of ~6.5 % to the $a$ direction ($a$=7.80 Å at strain-free constant $b$=12.69 Å) is needed to trigger the positive-to-negative transition of $J_{1a}$ (Fig. S2) while that for $J_{1b}$ of ~1.0 % along the $b$ direction ($b$=12.81 Å at strain-free constant $a$=7.33 Å) remains comparable with the value for $CrSe_2$.

Recently synthesized 2D $CrSe_2$ and $CrTe_2$ layers on various substrates and their magnetic characterizations [12, 13, 16] help with verifying the dependence of magnetic orders on lattice constants. In a $CrSe_2$ monolayer grown on $WSe_2$ [13], its lattice constant of 3.63 Å ($a$ = 7.26 Å and $b$ = 12.57 Å, black triangle in Fig. 2a)) sits in the FM region of our phase diagram, consistent with the weak FM behavior found in Ref. 9. A smaller lattice constant of 3.3 Å, corresponding to $a$=6.6 Å and $b$=11.4 Å (black star in Fig. 2a), was reported in another $CrSe_2$ monolayer grown on highly oriented pyrolytic graphite (HOPG). Its lattice constants reside in the ZZ-AFM region, responsible to the absence of ferromagnetic signals in x-ray magnetic circular dichroism (XMCD) measurements [16]. In terms of $CrTe_2$ monolayers, a sample prepared on a SiC-supporting bilayer graphene substrate shows lattice constants $a$ = 6.8 Å and $b$ = 12.15 Å (black star in Fig. 2b), locating in the ZZ region of the phase diagram, which was proved using spin-polarized scanning tunneling microscopy (SPSTM) measurements [12].

Interlayer magnetic coupling introduces additional complexity of magnetism in $CrX_2$ bilayers. Tunability of intralayer magnetism aside, it would be a more interesting and yet to be answered question that whether in-plane strain could tune interlayer magnetism in $CrX_2$ bi- or few-layers (Fig. 3a). In other words, we are interested if the varying in-plane lattice constants change out-of-plane magnetic coupling in $CrX_2$ bilayers, which was discussed in details as follows.

Figure 3b shows the phase diagram for the magnetic groundstate of the $CrSe_2$ bilayer over in-plane lattice constants $a$ and $b$. It is qualitatively comparable with that of the $CrSe_2$ monolayer in terms of intra-layer magnetism. Intralayer FM and interlayer AFM, (FM-AFM, $a$=7.10 Å and $b$=12.29 Å, see Fig. 3a) was used as the strain-free groundstate and thus the reference to calculate exact strain values in the $CrSe_2$ bilayer, different from that use for the monolayer (see Table S2 for more details). The easy axis in the bilayer only rotates 10° towards the z axis from that of monolayer. Equilibrium in-plane lattice constants of the $CrSe_2$ bilayer exhibit substantially enlarged in-plane lattice constants (blue dot in Fig. 3b) in comparison with that of the monolayer, as theoretically revealed in $CrS_2$ [17], $CrSe_2$ [13, 14] and $CrTe_2$ [12, 14] bilayers. It thus leads to the FM groundstate for the in-plane magnetism in the $CrSe_2$ bilayer.

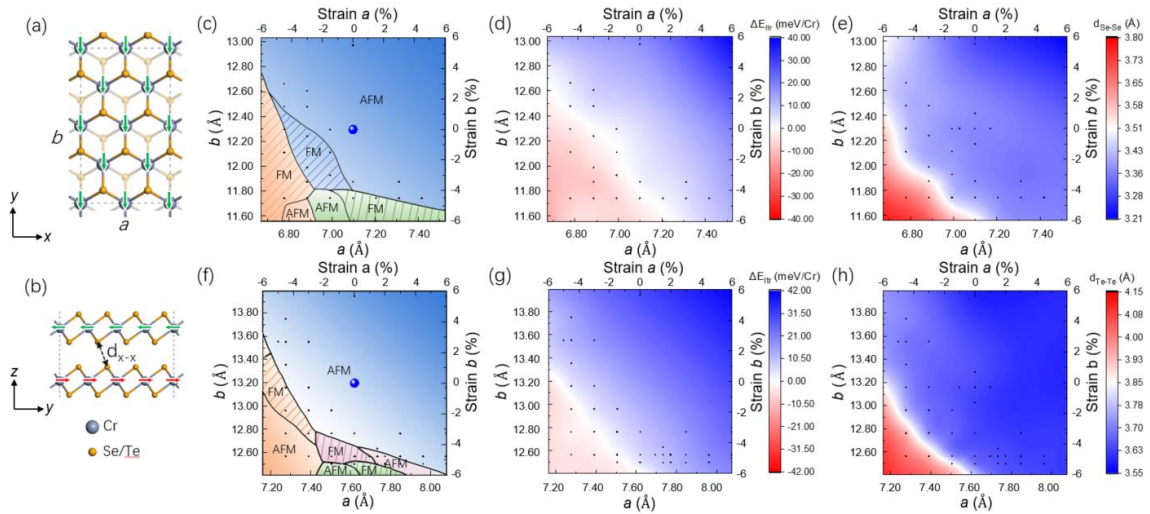

**Figure 3. Tunability of intra- and inter-layer magnetisms of $CrSe_2$ and $CrTe_2$ bilayers.** (a) and (b) Top and side views and magnetic ground state of an AA stacked $CrX_2$ bilayer. Grey and orange balls represent Cr and Se (Te) atoms, respectively. Lattice constants $a$ and $b$ are denoted in (a). Interlayer Se-

Se (Te-Te) distance is labeled by the black dash arrows in (b). Colored solid arrows on Cr represent the majority spin of up (red) and down (green). (c) Phase diagram of the $CrSe_2$ bilayer as a function of lattice constants $a$ and $b$. While the color scheme and presentation style of data points are as the same as that used for the diagrams of monolayers, those shadowed and plain colored regions indicate interlayer magnetic couplings are FM and AFM, respectively. Panel (d) plots the energy difference between interlayer FM and AFM states of the intra-layer FM configuration at different lattice constants. The blue (positive) region represents favored interlayer AFM state and FM for the red (negative) region. (e) Map of the interlayer Se-Se distance with respect to different $a$ and $b$ constants. Panels (f) to (h) duplicate panels (c) to (e) for the $CrTe_2$ bilayer case.

The interlayer magnetism in the bi- and thicker-layers shows fruitful tunability under in-plane strain. As shown in the phase diagram (Fig. 3c), each colored region (intra-layer magnetism), splits into two sub-regions denoting inter-layer FM (shadowed color) and AFM (plain color) configurations, respectively. The interlayer AFM-FM transition follows two rules depending on the intra-layer magnetism. The inter-layer AFM is more favored with larger in-plane lattice constants in the case of the intra-layer FM configuration. For example, the FM-AFM state of $CrSe_2$ undergoes an inter-layer magnetic transition to the FM-FM state under a 2.8% compressive strain applied in the $a$ direction. Here, we defined the energy difference between the FM-FM and FM-AFM states as $\Delta E_{Itr}$ and plotted its values for $CrSe_2$ in Fig. 3d where it shows strong in-plane strain dependence. The AFM interlayer exchange (positive $\Delta E_{Itr}$) is gradually suppressed by applying compressive strain to the either lattice. Here, we define the negative of ratio of the interlayer strain to the uniaxial in-plane strain as the interlayer Poisson's ratio. If a finite positive value of it is presented, the shrank in-plane lattice constants should enlarged inter-layer distance $d_{X-X}$ and thus more favor the inter-layer FM coupling, following the Bethe-Slater curve-like (BSC-like) behavior and the super-orbital mediated super-exchange mechanism that we previously revealed in $MX_2$ bilayers [14].

Figure 3e shows that $d_{Se-Se}$ could vary from 3.21 to 3.80 Å in the range of ±6 % in-plane strain for the $CrSe_2$ bilayer, which crosses the critical distance of 3.45 Å for the interlayer FM to AFM transition. The interlayer Poisson's ratio is 0.48 (0.46) under the uniaxial strain along the $a$ ($b$) direction for the $CrSe_2$ bilayer. Such significant response of the interlayer distance to the in-plane strain ensures the feasibility of tuning out-plane interlayer magnetic configuration by applying in-plane stress fields. An opposite trend of inter-layer magnetism and in-plane strain was found for the ZZ or ABAB intra-layer magnetism in the CrSe$_2$ bilayer, termed reversed BSC-like behavior, ascribed to a competition between super-orbital mediated direct- and super-exchanges, which will be elucidated elsewhere.

In terms of the CrTe$_2$ bilayer, its phase diagram (Fig. 3f) displays a comparable pattern with that of the CrSe$_2$ bilayer, except the missing FM-FM region and the additional intra-layer AABB configuration. While the interlayer magnetisms of the intra-layer ZZ-AFM and sAFM-ABAB configurations follow the reversed BSC-like behavior, the interlayer spin-exchange interaction in the intra-layer AABB configuration obeys the BSC-like behavior, i.e. interlayer AFM (FM) for larger tensile (compressive or smaller tensile) strains. Another distinct difference of the CrSe$_2$ and CrTe$_2$ cases lies in the missing FM-FM region in the CrTe$_2$ diagram. Energy $\Delta E_{\text{Itr}}$ of the CrTe$_2$ bilayer (Fig. 3g) flips its sign in regions where the preferred in-plane magnetism already transforms into ZZ-AFM, sAFM-AABB or sAFM-ABAB. Such delayed sign reversal is ascribed to a much larger transition distance of the CrTe$_2$ bilayer. Under the ±6% in-plane strain, interlayer distance $d_{\text{Te-Te}}$ of the CrTe$_2$ bilayer varies from 3.55 to 4.15 Å under, and the interlayer Poisson's ratio is around 0.61 (0.23) under the uniaxial strain along the $a$ ($b$) direction. Such distance range does not include the interlayer FM-AFM transition distance of 4.32 Å, but other in-plane magnetic configurations emerge before the interlayer AFM-FM transition occurs.

The phase diagrams imply that inter- and intra-layer magnetic orders could be tuned by in-plane strain engineering, which usually utilizes using slightly lattice-mismatched vdW substrates, namely the epitaxy strain. We thus considered an example, i.e. CrTe$_2$/bilayer graphene (BLG), to see how in-plane epitaxy strain behaves in determining its magnetism. Figure 4a depicts a schematic model of a $10 \times 3\sqrt{3}$ CrTe$_2$/$16 \times 4\sqrt{3}$ BLG superlattice, as experimentally determined in a previous work [12]. While the graphene-CrTe$_2$ stacking order varies from site to site within a domain (red hexagons), no apparent charge transfer and interlayer wavefunction overlap [12-14, 17] were observed between the CrTe$_2$ layer and the graphene substrate (Fig. 4b). The in-plane epitaxy strain effect thus plays a dominant role in tuning the magnetism of the epitaxial layer. In this particular case, the nearest Cr-Cr distances are 3.42 and 3.70±0.02 Å in the *a* and *b* directions (Fig. 4c left), respectively, which are within the ZZ region of the phase diagram and far from the phase boundaries (see black star in Fig.2b), consistent with the robust ZZ AFM state as depicted in Fig. 4c right.

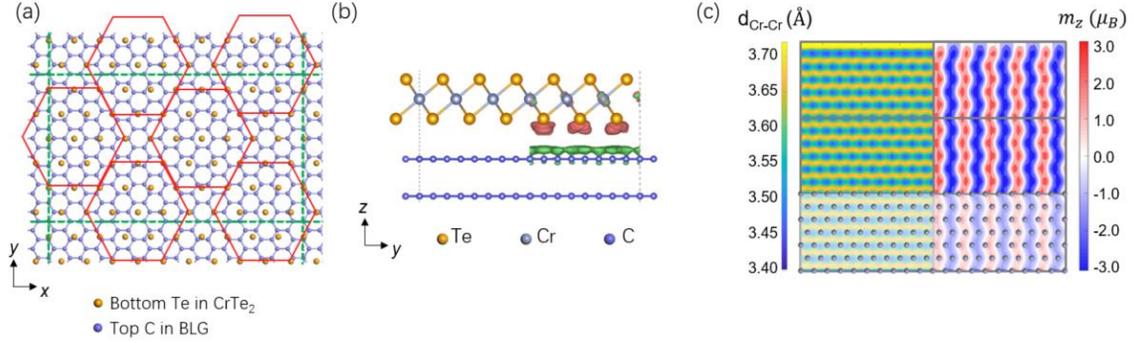

**Figure 4. Structure and magnetism of the epitaxial CrTe$_2$ monolayer on bi-layer graphene.** (a) Schematic model of a $10 \times 3\sqrt{3}$ CrTe$_2$/ $16 \times 4\sqrt{3}$ bilayer graphene (BLG) heterojunction. (b) Side view of the interlayer differential charge density contour of the heterojunction with an isosurface level of $2.0 \times 10^{-4}$ $e$ /Bohr$^3$. Red and green contours represent charge accumulation and depletion, respectively. (c) Mapping of the nearest Cr-Cr distance (left panel) and magnetic moments (right panel) of Cr atoms in the heterojunction.

In summary, we revealed that the in-plane epitaxy strain is capable of tuning both in-plane (intralayer) and out-of-plane (interlayer) magnetisms in epitaxy CrSe$_2$ and CrTe$_2$ mono- and bi-layers. In terms of intra-layer magnetism, the in-plane strain primarily changes the Cr-Se-Cr and Cr-Te-Cr angles that govern the intra-layer spin-exchange couplings, giving rise to the ZZ, ABAB and FM magnetic configurations. Our predicted phase diagrams of monolayers were confirmed with experimentally synthesized 2D CrSe$_2$ or CrTe$_2$ samples. A more striking effect lies in that the in-plane strain, through a finite Poisson's ratio, varies the inter-layer distance, which subsequently determines the inter-layer magnetism following a Bethe-Slater curve (BSC) like or a reversed BSC-like behavior depending on the in-plane magnetism. This exceptional effect enables of tuning out-of-plane magnetism using an in-plane strain field. Upon epitaxy of those 2D magnets on a graphene substrate, epitaxy strain still dominantly determines their magnetism because of suppressed interlayer charge transfer. Our calculation suggests a considerable magneto-elastic effect in 2D CrSe$_2$ and CrTe$_2$, and indicate that magnetisms of them can be manipulated in vdW epitaxy by in-plane strain from appropriately selected vdW substrates.

**Method**

DFT calculations are carried out using the generalized gradient approximation for the exchange-correlation potential, the projector augmented wave method [18] and a plane-wave basis set as implemented in the Vienna ab-initio simulation package (VASP) [19]. A $2 \times 2\sqrt{3}$ supercell of monolayer and bilayer CrX$_2$ is adopted to take consider of four intralayer magnetic configurations listed in Fig. 1(b)-(e) and a 10×6×1 k-mesh was used to sample the first Brillouin zone of it. A sufficiently large vacuum layer over 16 Å along the out-of-plane direction was adopted to eliminate the interaction among layers. The kinetic energy cut-off for the plane-wave basis was set to 700 eV for the geometric properties and 600 eV for electronic structure and energy calculations. Dispersion corrections were made at the van der Waals density functional (vdW-DF) level [20], with the optB86b functional for the exchange potential (optB86b-vdW) [21], which was proved to be accurate in describing the structural properties of layered materials [22-24] and was adopted for structure optimization. The shape and volume of mono- and bi-layer structures were fully relaxed until the residual force per atom was less than 0.002 eV/Å. A second-order Methfessel-Paxton smearing method with a sigma value of 0.01 eV was adopted during calculations. For energy comparisons among different magnetic configurations, we used the Perdew–Burke–Ernzerhof (PBE) functional [25] based on the vdW-DF-revealed structures. In-plane strain was defined as $\varepsilon = \frac{a-a_0}{a_0} \times 100\%$, where $a_0$ and $a$ denote lattice constants before and after applying in-pane strain. From the lattice constants of intralayer FM structure, which maintains the C3 symmetry, ±8% (±6%) strain was considered in both directions and one date point for each 1% (1.5%) strain variation was collected to calculate the phase diagrams of monolayers (bilayers). Besides, 0.5% strain for each point around phase transformation points is adopted to get phase boundaries. The interlayer Poisson's ratio is defined as $\nu_\perp = \frac{\Delta h/h}{\varepsilon_{11}+\varepsilon_{22}}$, where $\varepsilon_{11}$ and $\varepsilon_{22}$ is the in-plane strains along the a and b directions, and $h$ is the interlayer vertical Se-Se (Te-Te) distance (Fig. S5). To approximately estimate the interlayer Poisson's ratio, we made a linear fit on the variations of interlayer distances under a series of uniaxial strains in directions *a* or *b* in a range of $\pm1.5\%$ with a 0.5% step size (see Fig. S5 for more details). To account for on-site Coulomb interaction to the Cr d orbitals, U and J values of 4.5 eV (3.0 eV) and 0.6 eV, revealed by a linear response method [26] and by comparison with the results of HSE06 [27] in previous work [14], are adopted in $CrSe_2$ ($CrTe_2$).

**Acknowledgements**

We gratefully acknowledge financial support from the Ministry of Science and Technology (MOST) of China (Grant No. 2018YFE0202700), the National Natural Science Foundation of China (Grants No. 61761166009, No. 11974422 and No. 12104504), and the Strategic Priority Research Program of Chinese Academy of Sciences (Grant No. XDB30000000). C.W. was supported by the China Postdoctoral Science Foundation (2021M693479). L.W. was supported by the Outstanding Innovative Talents Cultivation Funded Programs 2021 of Renmin University of China. Calculations were performed at the Physics Lab of High-Performance Computing of Renmin University of China, Shanghai Supercomputer Center.

# Supplemental Information

# In-plane epitaxy strain tuning intralayer and interlayer magnetic couplings in CrSe$_2$ and CrTe$_2$ mono- and bi-layers

Linlu Wu[1, †], Linwei Zhou[1, †], Xieyu Zhou[1], Cong Wang[1, *] and Wei Ji[1, *]

[1]*Beijing Key Laboratory of Optoelectronic Functional Materials & Micro-Nano Devices, Department of Physics, Renmin University of China, Beijing 100872, P.R. China*

Corresponding authors: C.W. (email: wcphys@ruc.edu.cn), W.J. (email: wji@ruc.edu.cn)

† These authors contributed equally to this work.

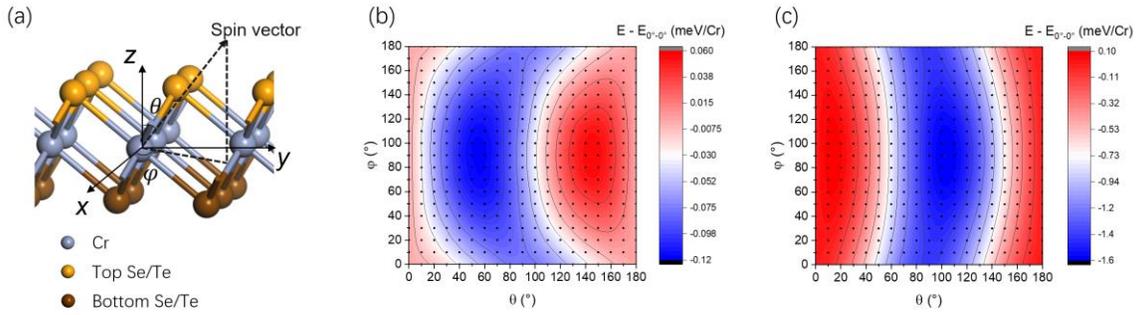

**Figure S1. Calculation of magnetic anisotropy energy (MAE) and easy axes of monolayer CrSe$_2$ and CrTe$_2$.** (a) Schematic model of monolayer CrX$_2$. The grey, orange and brown balls represent Cr, top-Se (Te) and bottom-Se (Te), respectively. Directions $x$, $y$, and $z$ correspond to those three lattice vectors and are perpendicular to each other. Angles θ and φ are defined to describe directions of spin vectors. (b) and (c) Calculation of MAE of monolayer CrSe$_2$ and CrTe$_2$, respectively. Relative total energies with magnetization directions fixed in different directions (with respect to $E_{0°-0°}$, which denotes the energy with magnetization directions fixed in the $z$ direction). Easy axes are in the $y$-$z$ plane and are 60° and 100° off the $z$ axis in CrSe$_2$ and CrTe$_2$, respectively.

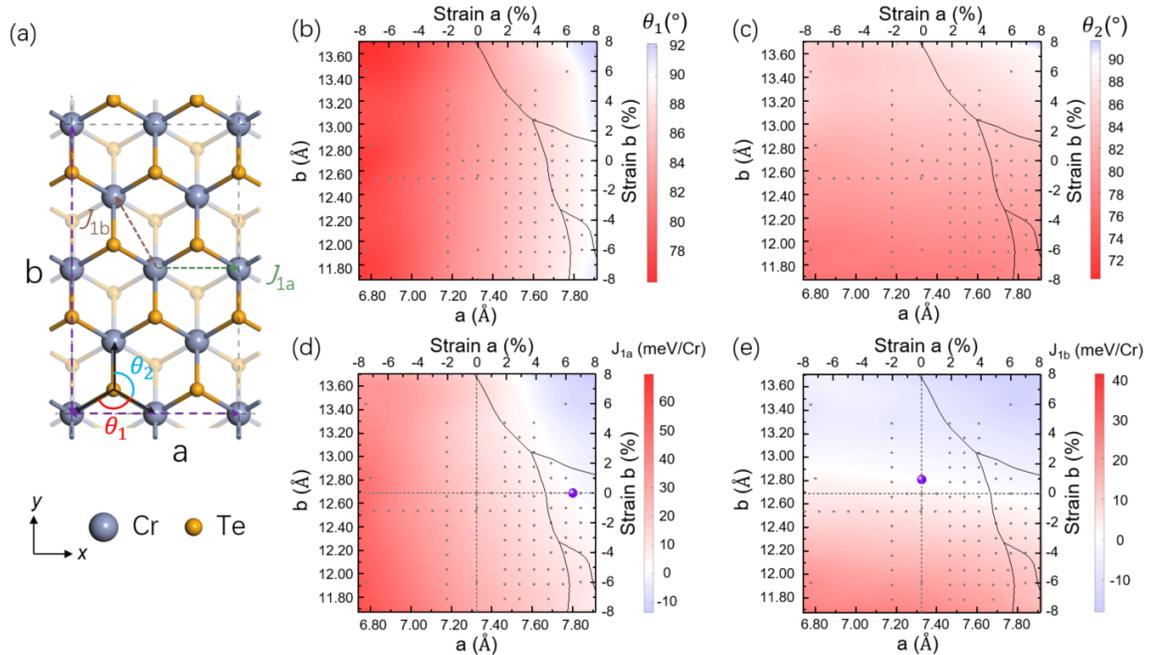

**Figure S2. Variations of Cr-Te-Cr angles and exchange parameters with lattice constants in monolayer CrTe$_2$.** (a) Top view of a $2 \times 2\sqrt{3}$ supercell of monolayer CrTe$_2$. Lattice constants $a$ and $b$ are labeled by purple dash lines. The Cr-Te-Cr angles $\theta_1$ and $\theta_2$ are labeled by arcs. Exchange parameters $J_{1a}$ and $J_{1b}$ are labeled by green and brown dash lines, respectively. (b) and (c) $\theta_1$ and $\theta_2$ vary with lattice constants in monolayer CrTe$_2$. (d) and (e) Mapping of exchange parameter $J_{1a}$ and $J_{1b}$

as a function of lattice constants. Positive (red) region and negative (blue) region represent AFM and FM exchange, respectively. Phase boundaries of monolayer CrTe$_2$ are labeled by black lines in (b) - (e).

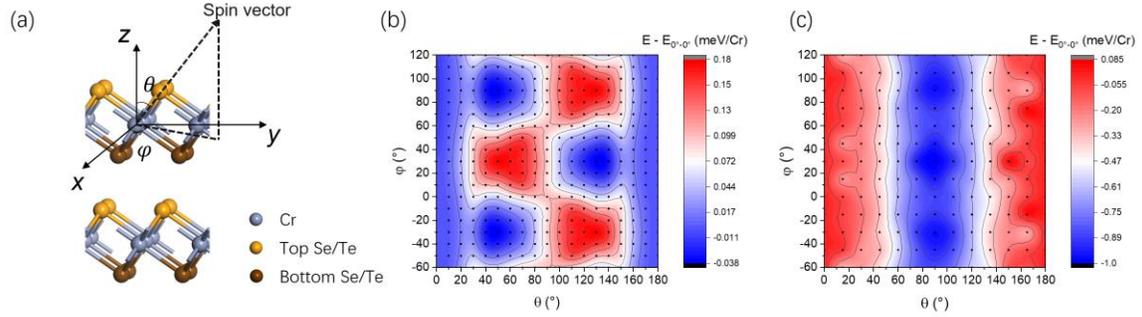

**Figure S3. Calculation of magnetic anisotropy energy (MAE) and easy axes of bilayer CrSe$_2$ and CrTe$_2$.** (a) Schematic model of bilayer CrX$_2$ (X=Se, Te). The grey, orange and brown balls represent Cr, top-Se (Te) and bottom-Se (Te) in each layer, respectively. Directions *x*, *y*, and *z* correspond to those three lattice vectors and are perpendicular to each other. Angles θ and φ are defined to describe directions of spin vectors. (b) and (c) Calculation of MAE of bilayer CrSe$_2$ and CrTe$_2$, respectively. Relative total energies with magnetization directions fixed in different directions (with respect to $E_{0°-0°}$, which denotes the energy with magnetization directions fixed in the *z* direction). Easy axis of bilayer CrSe$_2$ is in the *y-z* plane and 50° off the *z* axis, while in bilayer CrTe$_2$ it is along the *y* axis.

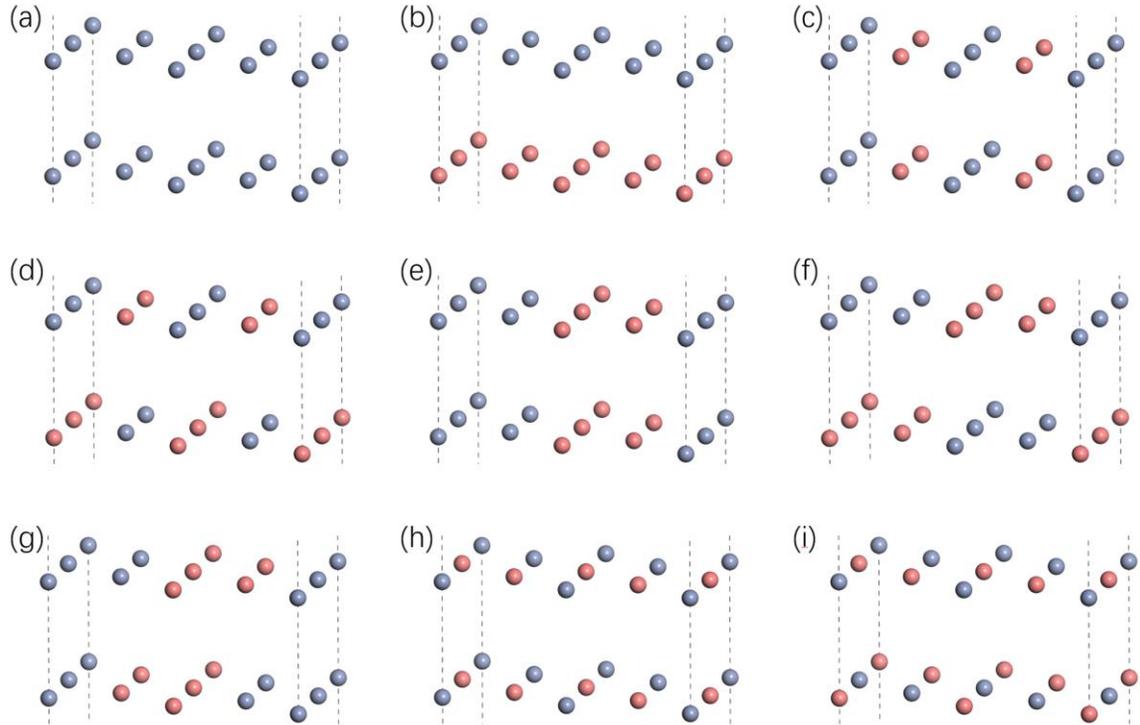

**Figure S4. Schematic representation of nine magnetic orders considered in bilayer CrSe$_2$ and CrTe$_2$.** The gray and pink balls represent the magnetic moment up and down on Cr atoms, respectively. (a) FM-FM; (b) FM-AFM; (c) ABAB-FM; (d) ABAB-AFM; (e) AABB-FM; (f) AABB-AFM1; (g)

AABB-AFM2; (h) ZZ-FM; (i) ZZ-AFM. The first and second labels represent intralayer and interlayer magnetism, respectively.

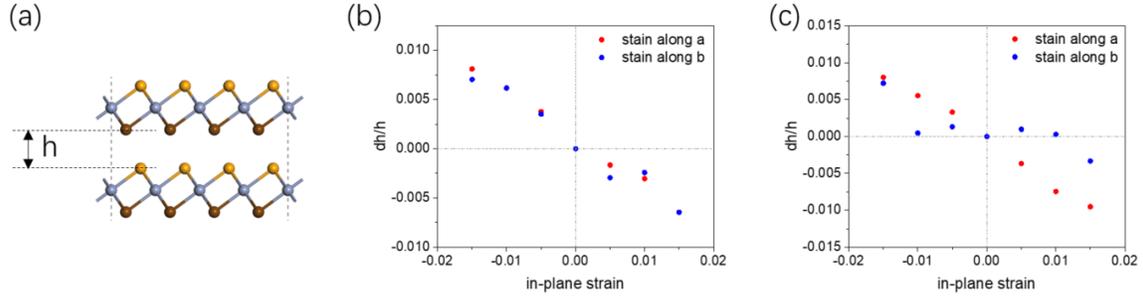

**Figure S5. Calculation of interlayer Poison's ratio in bilayer CrSe$_2$ and CrTe$_2$.** The interlayer Poisson's ratio is defined as $v_\perp = \frac{\Delta h/h}{\varepsilon_{11}+\varepsilon_{22}}$, where $\varepsilon_{11}$ and $\varepsilon_{22}$ is the in-plane strains along the a and b directions, and $h$ is the interlayer vertical Se-Se (Te-Te) distance. (a) Side view of bilayer CrSe$_2$ (CrTe$_2$) and definition of interlayer vertical distance. (b) and (c) Variations of $\Delta h/h$ on uniaxial strain along the $a$ ($b$) in bilayer CrSe$_2$ and CrTe$_2$, respectively.

**Table S1.** Geometric and magnetic details of monolayer CrSe$_2$ and CrTe$_2$. Four magnetic configurations are labeled in Fig 1(b)-(d). 'Mag. Config.' is the abbreviation of magnetic configuration. $\triangle$E is the energy difference from the ground state. Nearest Cr-Cr distances are listed in the table. $\theta_1$, $\theta_2$ and $\theta_3$ are labeled in Fig. 1a. Structures are optimized with optB86b-vdW+UJ, and all energy and magnetic moment related values are calculated using PBE+UJ.

| Fully relaxed | Intralayer Mag. Config. | $\triangle$E (meV/Cr) | a (Å) | b (Å) | Cr-Cr distance (Å) | | | Angle of Cr-X-Cr (°) | | | Mag. Mom. (μB) | |
|---|---|---|---|---|---|---|---|---|---|---|---|---|
| | | | | | $r_1$ | $r_2$ | $r_3$ | $\theta_1$ | $\theta_2$ | $\theta_3$ | Cr | Se |
| ML-CrSe$_2$ | FM | 22.40 | 6.84 | 11.84 | 3.42 | 3.42 | 3.42 | 84.6 | 84.6 | 84.6 | 3.09 | -0.21 |
| | sAFM-ABAB | 0.00 | 7.00 | 11.26 | 3.50 | 3.32 | 3.32 | 86.8 | 81.7 | 81.7 | 3.02 | 0.04 |
| | sAFM-AABB | 10.08 | 6.96 | 11.43 | 3.48 | 3.38 | 3.38 | 86.0 | 83.4 | 83.4 | 3.12 | 0.18/0.06 |
| | ZZ | 2.97 | 6.75 | 11.68 | 3.38 | 3.42 | 3.34 | 83.4 | 82.3 | 84.5 | 3.03 | 0.03 |
| ML-CrTe$_2$ | FM | 39.57 | 7.33 | 12.66 | 3.66 | 3.66 | 3.66 | 84.3 | 84.1 | 84.1 | 3.06 | -0.18 |
| | sAFM-ABAB | 3.25 | 7.52 | 11.98 | 3.76 | 3.54 | 3.54 | 86.5 | 80.8 | 80.8 | 3.02 | 0.04 |
| | sAFM-AABB | 5.49 | 7.50 | 12.12 | 3.75 | 3.63 | 3.63 | 86.0 | 83.1 | 83.1 | 3.14 | 0.16/0.06 |
| | ZZ | 0.00 | 7.18 | 12.54 | 3.59 | 3.66 | 3.57 | 82.1 | 81.7 | 83.9 | 3.06 | 0.04 |

**Table S2.** Energy comparation of nine magnetic configurations in bilayer CrSe$_2$ and CrTe$_2$. Structures are optimized with optB86b-vdW+UJ. In-plane FM is kept in structure optimization. Whether interlayer AFM or FM are adopted in structure optimization, the energetically favored magnetism is always FM-AFM (intralayer FM and interlayer AFM). Each row in the table lists relative energy of nine magnetic orders with FM-AFM.

| Unit (meV/Cr) | Fully relaxed structure | Mag. Config. | | | | | | | | |
|---|---|---|---|---|---|---|---|---|---|---|
| | | FM-FM | FM-AFM | ABAB-FM | ABAB-AFM | AABB-FM | AABB-AFM1 | AABB-AFM2 | ZZ-FM | ZZ-AFM |
| BL-CrSe$_2$ | FM-AFM | 2.42 | 0.00 | 31.42 | 33.73 | 21.41 | 28.21 | 26.71 | 31.39 | 32.76 |
| | FM-FM | 1.21 | 0.00 | 31.27 | 33.49 | 20.69 | 27.54 | 25.71 | 28.22 | 29.77 |
| BL-CrTe$_2$ | FM-AFM | 7.47 | 0.00 | 40.21 | 43.38 | 22.53 | 29.11 | 22.24 | 36.44 | 36.65 |
| | FM-FM | 6.64 | 0.00 | 38.34 | 41.45 | 21.29 | 27.67 | 21.08 | 34.02 | 34.49 |

**Table S3.** Geometric and magnetic details of bilayer CrSe$_2$ and CrTe$_2$. 'Mag. Config.' is the abbreviation of magnetic configuration. Lattice constants '*a*' and '*b*' are labeled in Fig. 3a. '*r*' denotes the nearest Cr-Cr distance. Layer height is the vertical distance between the top and bottom Se (Te) sublayer in a Se-Cr-Se (Te-Cr-Te) layer. Interlayer Cr-Cr distance and Se-Se (Te-Te) distance (labeled in Fig.3a) are also listed in the table. △E is the energy difference from the ground state.

| | Mag. Config. | a (Å) | b (Å) | r (Å) | layer height (Å) | interlayer Cr-Cr (Å) | interlayer X-X (Å) | △E (meV/Cr) | mag Cr (μB) | mag Se (μB) |
|---|---|---|---|---|---|---|---|---|---|---|
| BL-CrSe$_2$ | FM-AFM | 7.10 | 12.29 | 3.55 | 3.00 | 5.67 | 3.37 | 0.00 | 3.14 | 0.21 |
| | FM-FM | 7.06 | 12.23 | 3.53 | 3.01 | 5.71 | 3.39 | 2.40 | 3.10 | -0.20 |
| BL-CrTe$_2$ | FM-AFM | 7.62 | 13.20 | 3.81 | 3.22 | 6.11 | 3.64 | 0.00 | 3.14 | 0.18 |
| | FM-FM | 7.60 | 13.17 | 3.80 | 3.22 | 6.15 | 3.66 | 7.47 | 3.09 | -0.17 |